\documentclass[%
 reprint,
superscriptaddress,
showpacs,preprintnumbers,
amsmath,amssymb,
aps,
prper,
]{revtex4-2}

\usepackage{graphicx}
\usepackage{dcolumn}
\usepackage{bm}
\usepackage{hyperref}
\usepackage{longtable}

\begin{document}

\title{Modern physics courses: Understanding the content taught in the U.S.}

\author{Alexis Buzzell}
\author{Ram\'on Barthelemy}
\affiliation{%
 Department of Physics \& Astronomy, University of Utah
}%

\author{Tim Atherton}%
\affiliation{%
 Department of Physics \& Astronomy, Tufts University
}%

\begin{abstract}
  The modern physics course is a crucial gateway for physics majors, introducing concepts beyond the scope of K-12 education. Despite its significance, content varies widely among institutions. This study analyzes 167 modern physics syllabi from 127 US research intensive institutions, employing emergent coding using both human and Natural Language Processing methods from public sources (51.5\%) and private correspondence (48.5\%). Public course catalogs were consulted to identify pre- and co-requisites, with 37.1\% of students having completed calculus II. Foundational topics like Newtonian mechanics (94\%), electricity and magnetism (84.4\%), and waves or optics (77.2\%) were frequently required. Quantum physics (94\%), atomic physics (83\%), and relativity (70\%) were most commonly taught. The study highlights the lack of uniformity in modern physics curricula, emphasizing the importance of a consistent and comprehensive education for physics majors across universities. This insight contributes to the ongoing discourse on optimizing physics education in higher education.
\end{abstract}

\maketitle

\section{Introduction}
The modern physics course serves as a pivotal gateway for students pursuing a physics major, introducing them to new material early in their undergraduate coursework beyond the scope of their K-12 experience \cite{Knight}. Despite its significance, there exists considerable variability in the topics covered, both across different institutions and even within multiple offerings of modern physics at the same institution. 

The lack of uniformity in modern physics curricula poses a challenge, impacting the consistency of education received by physics majors. Achieving uniformity across institutions is crucial to ensuring that students receive a standardized and comprehensive education, adequately preparing them for advanced studies and future careers, regardless of their undergraduate institution. This paper presents a comprehensive analysis using both human and Natural Language Processing (NLP) methods of 167 modern physics syllabi from 127 research intensive institutions, and identifies the most commonly taught topics across the US. 

\section{Literature review}
\subsection{Education research on the modern physics course}
Discussions around the need for better alignment between modern physics courses and the more advanced quantum mechanics courses have been taking place since 2001 \cite{Singh2001}; Singh argued the semi-classical models learned in modern physics courses can give rise to the misconceptions students must overcome in their quantum mechanics courses. Modern physics courses need to place an emphasis on the limits and appropriate applications for the semi-classical models learned. Vokos et al. \cite{Vokos2000}, stated that an instructional goal for modern physics is for students to be introduced to quantum concepts at the level of understanding wave-particle duality. This message is echoed by Singh et al. \cite{Singh2006}, by stating most physicists were introduced to quantum mechanics during their modern physics courses as undergraduate students. However, neither statements provided empirical evidence of this claim; hence the need for this study arose to determine what topics are taught in modern physics courses within the US, and is this students' first exposure to quantum concepts in their curriculum. 

Other research on modern physics has predominantly focused on students' misconceptions and the development of their understanding of quantum mechanics. Researchers have explored students' ontological and epistemological shifts as they transition from a classical physics perspective to a quantum perspective \cite{Baily2009, Hadzidaki2000}. Investigations have also delved into challenges related  to learning quantum tunneling \cite{McKagan2008} and the development of assessment tools to evaluate conceptual and visual understanding of quantum mechanics \cite{Cataloglu2002}. 

In response to the diverse needs of students, particularly engineering majors, reforms have been implemented to tailor modern physics curricula, emphasizing real-world applications over abstract problems \cite{McKagan2007}. Arguments have also been made for introducing modern physics topics earlier in physics education, either at the high school or lower undergraduate levels \cite{Aubrecht1986, Michelini2016}. 

Some scholars have focused on instructor perspectives on the modern physics course. An informal survey conducted via The Physics Teacher (December 2013 edition) and the 2014 Oersted Lecture sought opinions on essential introductory modern physics topics \cite{Zollman2016}. Quantum mechanics and special relativity were considered crucial, while thermodynamics and rotational dynamics were suggested for omission. This current study builds off Zollman's informal survey by providing a detailed analysis of the topics taught in modern physics courses. 

\subsection{Literature on STEM syllabi}
When considering how best to determine the topics taught across modern physics courses, syllabi readily became a feasible option. Syllabi served as a valuable tool to obtain course content without significant time commitments from the course instructor. However, they also reflect instructors' and institutions values within STEM \cite{Parson2016}. Syllabi, beyond structuring learning outcomes and course objectives, communicate expectations between instructors and students. Epistemological beliefs embedded in syllabi impact pedagogical approaches. The language used in syllabi is crucial, as studies have shown that a "chilly climate," in the classroom, characterized by male-normed, highly impersonal, and individualistic environment, can lead to women changing their major to non-STEM fields. 

\subsection{Artificial intelligence methods in PER}
To complement the human-coded analysis, this study also performed analysis of the collected data using Artificial Intelligence (AI) and NLP techniques. Driven by availability of new technologies for automated processing of large textual datasets, there has been an emerging interest in applying such methods to data collected in PER studies and in higher education more broadly \cite{zawacki2019systematic}. Many algorithms are available to solve a variety of tasks\cite{jurafsky2009speech, mitkov2021}, falling broadly into \emph{supervised} techniques where a set of training data must be provided to the algorithm and \emph{unsupervised} techniques that discover structure without human supervision. To situate our work, we provide a brief snapshot of how NLP methods are already being applied in PER before turning to our analysis.

One possible application domain is to analyze student work. Wilson \emph{et al.} created classifiers for free responses to the Physics Measurement Questionnaire\cite{PhysRevPhysEducRes.18.010141} that performed with similar reliability as two human coders, but urged caution in using such approaches in the classroom due to the possibility of biases in the classifier or in the training set. Similarly, Campbell \emph{et al.} \cite{PhysRevPhysEducRes.20.010116} used the Watson NLP to classify whether certain conceptual themes were present or absent in short-answer student responses.

Another NLP task is to identify latent themes within a textual corpus. Odden and coworkers \cite{PhysRevPhysEducRes.16.010142} examined the contents of all PER Conference proceedings, some 1300 short papers, between 2001 and 2018 to resolve emergent topics and how they appeared in time; the same authors have further refined their method and\cite{odden2024using}. Such methods may be particularly appealling to scale qualitative research studies, in particular, because these tend to generate large quantities of text through transcripts or survey responses that can be time-consuming to analyze. Tschisgale \emph{et al.} recently advocate for incorporating NLP methods to facilitate theory-building in qualitative studies\cite{PhysRevPhysEducRes.19.020123}. 

A third possibility is to use AI tools to generate or answer physics problems. Large Language Models (LLMs), notably ChatGPT, have attracted considerable public and public attention because they generate naturalistic answers to questions posed in human language. Multiple studies have found that ChatGPT is capable of generating convincing answers to the Force Concept Inventory\cite{PhysRevPhysEducRes.19.010132,kieser2023educational}. However, Dahlkemper \emph{et al.} found that students could distinguish ChatGPT generated answers to physics problems from instructor-generated answers, but only if their subject knowledge was adequate\cite{PhysRevPhysEducRes.19.010142}. LLMs have also been demonstrated as part of a system intended to grade student work\cite{PhysRevPhysEducRes.19.020163}. Use of NLP in such contexts nonetheless raises a number of important ethical issues, due to biases in training sets and algorithmic features \cite{Gouvea}.

While AI methods are relatively new to PER, the corpus of syllabi obtained for the present study presents an attractive target for them: it is large enough to provide significant insight into the curricular content of a course that is taught in almost all physics departments, but small enough that human analysis remains feasible. Applying NLP to curricular-level research, as opposed to student work or understanding PER itself, appears to be a new application of these methods and therefore may become a useful tool for further studies. Like Odden et al. \cite{PhysRevPhysEducRes.16.010142}, we also believe that \emph{``{[}NLP{]} cannot replace careful analysis by humans''} and that \emph{``validation is very important''} and hence we chose to perform a multimethods approach centralizing human coding but supplementing and validating this with an exploratory NLP analysis. 

\section{Methods}
\subsection{Human-coding methods}
Syllabi	were collected from the institutions listed on US News Rankings of "The Best Physics Programs" \cite{USNEWS}. Of the 190 programs listed in the ranking, 181 offered a modern physics or equivalent course in their publicly available course catalog. From the 181 programs, syllabi were collected from 70.2\%, resulting in 127 institutions represented in the data set. Some programs offered more than one modern physics course to their student body, resulting in a total of 167 syllabi obtained from the 127 institutions. 

Within the set of 127 institutions, 78\% were classified as very high research activity, 20\% were classified as high research activity, and 2 institutions were not classified in the Carnegie Classification of Institutions of Higher Education \cite{CarnegieClassification}. 73\% are public institutions \cite{CarnegieClassification}. The syllabi were collected using public online searches (51.5\%) and private email communications with instructors and department administrative staff (48.5\%). 

The syllabi and publicly available course descriptions were analyzed to determine: (1) the content taught, (2) the prerequisites or corequisites to enroll, (3) the major the course is intended for, (4) the academic year in the four-year program during which students are anticipated to enroll, (5) instructors pedagogical approach, (6) grading scheme utilized, (7) policies listed. 

\textit{Content taught:}

An iterative emergent coding method was used to develop the codes for each topic \cite{Creswell} Table (\ref{tab:Topicscoding}) in the Appendix. The final coding scheme encompassed the following topics: (1) thermal physics, (2) relativity, (3) quantum mechanics, (4) atomic physics, (5) nuclear physics, (6) molecular physics, (7) solid state physics, (8) statistical physics, (9) cosmology, (10) programming skills, (11) mathematical foundations, (12) history of modern physics, (13) particle physics, (14) waves, optics, lasers and/or light, (15) astrophysics, (16) Lagrangian or Hamiltonian mechanics. In order for a course to be coded as including a topic, the syllabus has to list one or more of the codes for the topics listed in \ref{tab:Topicscoding}. A heat map was created from the topics taught as well. Using the counts for the individual topics, the correlation of topics taught together were found and mapped onto a heat map. 

\textit{Pre- and co-requisite requirements:}

Publicly available course catalogs were utilized to determine the highest level of mathematics and all physics topics required to enroll. The highest level of mathematics required to be taken prior or during the semester of enrollment in modern physics was found in the modern physics course description of the catalog. Courses were coded as requiring (1) no mathematics required, (2) Algebra, (3) Precalculus, (4) Calculus I, (5) Calculus II, (6) Calculus III, and (7) advanced mathematics. Any courses taken after calculus III, such as linear algebra or differential equations, were considered "advanced mathematics," as math courses are not necessarily taken in a linear progression after the completion of the calculus series. 

Physics topics required prior to enrollment were also found using the modern physics course description within the catalog. The prerequisite physics course code was recorded and then located in the course catalog. The course description was then utilized to code for each topic taught in the course. This step was necessary, as not all "Physics I" or "Physics II" courses encompass the same topics, while mechanics and electricity and magnetism are most commonly referred to by these course titles, other topics such as thermodynamics, waves, or special relativity may be included as well. The course description of the prerequisite course was used to discern this variability. The course description of the prerequisites' prerequisite course was then recorded and the course description used to code for topics taught, until no physics prerequisites were required. This iterative process allowed for coding of all physics topics required to enroll in the modern physics course. 

\textit{Intended major of students enrolled:}

The intended major of students enrolled in the modern physics course was coded for to determine the audience the institution intended the course to be tailored to. Using the course catalogs again, the physics degree requirements and sample four year timelines (if available), were referenced to determine if the modern physics course was intended specifically for physics majors and if the course was a requirement for graduation. 

To determine if the course was intended for majors other than physics students, the course description in the catalog was used. An example for this code would be if the course description included a statement such as "This course is intended for students majoring in physics, philosophy, mathematics, or engineering."  If no major was listed in the description as the audience, the assumed audience was physics students, as all courses in this study were listed within the institution's physics department. 

The physics degree requirements were also referenced to determine if the course was required for a physics major to graduate. While many of these courses intended audiences were physics majors specifically, the authors also recorded if the courses needed to be taken for the students to receive their four year degree. 

\textit{Year intended for enrollment:}

Multiple methods were used to code for which year (i.e. freshman, sophomore, junior, senior) students were expected to enroll in the course. Some courses included this information in the course description, in which case the course was coded for using this method. If the information was not available within the description, physics degree sample timelines were referenced when available. In the event neither of these methods were available, the physics prerequisite courses were used. If there were for example two required prerequisite courses to enroll in the modern physics course, it was then assumed students were expected to enroll in their third semester, or sophomore year. 

\textit{Pedagogical approach:}

Instructors' pedagogical approach was coded for using an emergent coding method \cite{Creswell}. Using an iterative process, the final codes were (1) lecture based, (2) lecture based supplemented with discussions, recitations, or in class activities, (3) not defined, (4) active classroom, (5) studio based, and (6) flipped or reverse classroom. The codes for each approach are listed in Table \ref{tab:pedagogycoding} in the Appendix. 

\textit{Grading scheme:}

The grading scheme used was coded for using an emergent coding method \cite{Creswell}. The codes for grading were (1) skills based, (2) curved, (3) may be curved, (4) may be curved but only for student benefit, (5) no curve, (6) not stated if there will be a curve or not, (7) pass/fail, (8) rounding policy stated for when students are on or close to a letter grade boundary. The codes for each grading scheme can be found in Table VI of the Appendix. 

Syllabi were also coded into three categories: (1) those that explicitly use exams as a form of assessment, (2) those that explicitly do not use exams as a form of assessment, (3) those that do not state whether exams are used or not as a form of assessment. The syllabi in category (1) those that explicitly use exams as a form of assessment were further divided into three subcategories: (a) those that explicitly have one or more cumulative exam, (b) those that explicitly have non-cumulative exams, (c) those that do not state if exam will be cumulative or non-cumulative. 

\textit{Policies:}

Policies were graded using an emergent coding method as well \cite{Creswell}. The final codes can be found in Table VII of the Appendix. The policies coded for included (1) academic integrity, (2) ADA accommodations, (3) FERPA, (4) religious observances, (5) exam policy, (6) late or makeup work, (7) EDI, sexual harassment, or Title IX statements, (8) basic needs resources, (9) attendance, (10) counseling services, (11) regrade policy, (12) email policy, (13) COVID-19, (14) academic success resources, (15) campus safety, (16) 2nd Amendment, (17) AI/Chat GPT usage, (18) classroom etiquette, (19) inclement weather, and (20) pregnancy or childbirth. 

\textit{Focus on institutions supporting diverse student bodies:}

In order to ensure the inclusion of institutions that serve the Black, Hispanic, and other diverse communities a separate analysis was conducted to look at the frequency rate of topics taught at Historically Black Colleges and Universities (HBCU, $n=1$), Hispanic Serving Institutions (HSI, $n=11$), Asian American and Native American Pacific Islander-Serving Institutions (AANAPISI, $n=10$), Predominantly Black Institutions (PBI, $n=1$), and Alaska Native-Serving Institutions or Native Hawaiian-Serving Institutions (ANNH, $n=1$) \cite{MSIDirectory}. 

\subsection{Topic modeling using NLP}
The analysis was framed as a \emph{topic modeling} NLP task\cite{vayansky2020review,mitkov2021,abdelrazek2023topic}. Topic modeling algorithms aim to learn \emph{topics} or hidden semantic patterns that exist in a corpus of text documents. They do so through a sequence of transformations: First the documents are \emph{tokenized}, converted to smaller units; the resulting tokens are then \emph{vectorized}, i.e. converted to a numerical representation; the algorithm then fits the encoded documents to discover topics from the vector representation. Each of these steps can utilize a number of subalgorithms. Additionally, topic modelling algorithms generally incorporate \emph{hyperparameters}, user selectable parameters that control the behavior of the algorithm. It is necessary as part of the analysis to conduct a human or automated
exploration of the topics identified as a function of these hyperparameters and perform an assessment of the quality of the identified topics. 

The corpus analyzed here comprised $n=169$ documents, largely in Adobe PDF format (151 files), 13 Microsoft Word files, 2 Microsoft Excel files, 1 HTML file, 1 plain text file and 1 PNG; all of these were analyzed except for the single png file. All files were converted to plain text for further analysis using the \texttt{pdftotext} utility for PDF files and \texttt{pandoc} for the remaining filetypes. As is typical in NLP methods, each plain text file was then \emph{cleaned} by converting all capitals to lower case; removing URLs; converting newlines, punctuation and control characters to spaces; and consolidating successive spaces. 

We first attempted to apply Latent Dirichlet Analysis to the corpus, which has previously been applied to perform a topic analysis of PERC proceedings \cite{PhysRevPhysEducRes.16.010142,odden2024using}.
Latent Dirichlet Analysis assumes that each document consists of a number of topics and that each token in a document is associated with one of the document's topics. It is necessary to remove commonly occurring
words or stopwords from the corpus prior to analysis. Despite exploring a wide range of hyperparameters, we did not find satisfactorily coherent topics. In part, this is likely due to the size of the corpus, which is much smaller than that explored in \cite{PhysRevPhysEducRes.16.010142,odden2024using}, but it is also due to limitations of the algorithm: For example, the order of the tokens isn't taken into account by LDA other than, optionally, as short sequences or \emph{n-grams} of tokens. We therefore turned to a newer class of algorithms that \emph{do} take token order into account and are pretrained on a much larger corpus. 

Behind the recent explosion of LLMs is the 2017 creation of the \emph{transformer} deep learning architecture, which is able to contextualize words within their surrounding environment, a \emph{context window} of a specified number of tokens. Using non-local information allows such models to better capture semantic structure, and the transformer architecture also facilitates parallelization for better performance. An early successful language model in this class, \texttt{BERT}\cite{devlin2018bert} (Bidirectional Encoder Representations from Transformers) remains an important baseline model for NLP tasks. Such models are pretrained on a corpus of data, in BERT's case English Wikipedia articles.

Here, we use the \texttt{BERTopic} topic modelling algorithm\cite{grootendorst2022bertopic} that performs the sequence: \emph{embedding} the corpus into a numerical representation using the \texttt{BERT} language model; \emph{dimensionality reduction} into a smaller parameter space; \emph{clustering} in the reduced space---this is the step that actually identifies the topics; and then building a \emph{representation} of the topics. The modular design means each component can be replaced as new submodels become available. Additionally, BERTopic provides a number of hyperparameters, but there is less need for tuning than earlier techniques. Importantly, the default clustering algorithm, \texttt{HDBSCAN}, automatically selects the number of clusters by finding a cluster size $\epsilon$ such that changes in $\epsilon$ do not change the number of clusters generated. Our analysis sequence is similar to that used in Ref. \cite{PhysRevPhysEducRes.19.020123}, although the underlying language model used here is necessarily different due to the English texts. 

In line with recommendations for usage, we modified our cleaning step to divide each text into fragments approximately corresponding to sentences using the \texttt{sent\_tokenize} function in the \texttt{nltk} package. Dividing the corpus into sentences, we obtained 11494 fragments in total with a mean length of 69 fragments per document. In contrast with LDA, it is not recommended to remove stopwords; the transformer architecture implicitly uses these words in understanding the context of other words.

\section{Results}
\textit{Content taught:}

\begin{figure*}
    \centering
    \includegraphics[width=400 pt, height=300 pt]{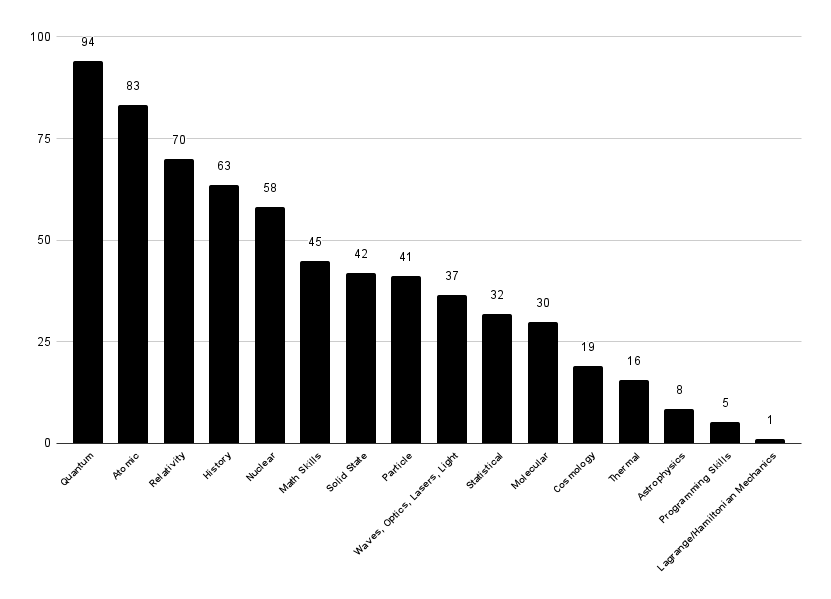}
    \caption{Topics taught in 167 modern physics courses across the US}
    \label{fig:enter-label}
    \includegraphics[width=400 pt, height=300 pt]{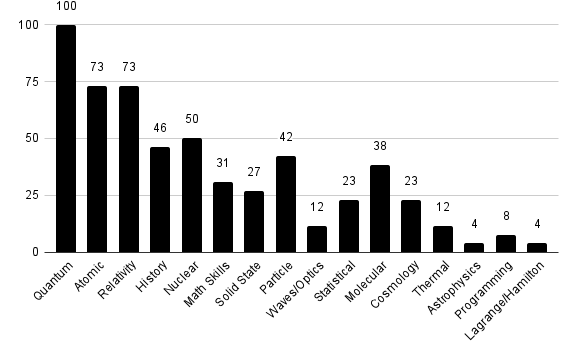}
    \caption{Topics taught in modern physics courses at MSIs}
    \label{fig:enter-label}
    
\end{figure*}
The results of the syllabi analysis concluded that quantum (94\%), atomic (83\%), and relativity (70\%) were the most commonly taught topics in modern physics within the US, as shown in Figure 1. Most courses also cover the historical background of modern physics (63\%). Figure 2 shows the distribution of content taught at MSIs represented in the data set. Within the MSIs, quantum was always taught (100\%), atomic was taught less often than the larger data set (73\%), and relativity was taught slightly more than the larger data set (73\%). 

\begin{figure}
    \centering
    \includegraphics[width=200 pt, height=200 pt]{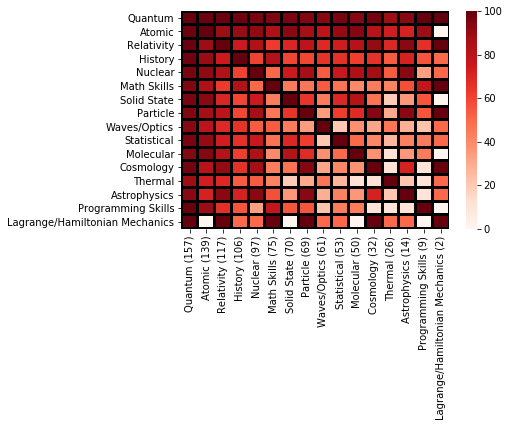}
    \caption{Correlation between topics taught}
    \label{fig:enter-label}
\end{figure}

Additionally a heat map, shown in Figure 3, was created to visually demonstrate the correlation of individual topics taught with other topics.  Figure 3 demonstrates when atomic topics were included in a syllabus, quantum was also included 98.6\% of the time. Whereas, when relativity was included, quantum was also listed 99.1\% of the time. 

\begin{figure*}
    \centering
    \includegraphics[width=400 pt, height=300 pt]{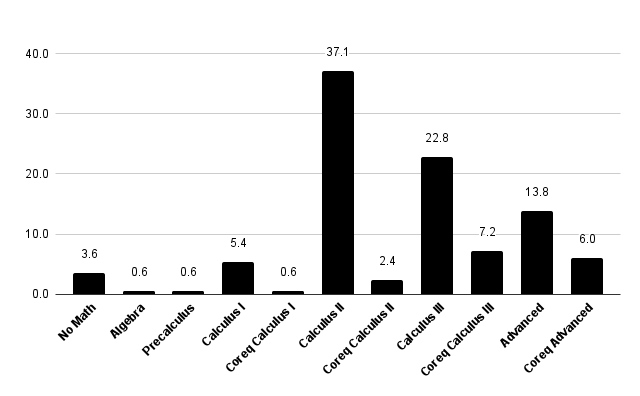}
    \caption{Mathematics pre- and co-requisites}
    \label{fig:enter-label}
    \includegraphics[width=400 pt, height=300 pt]{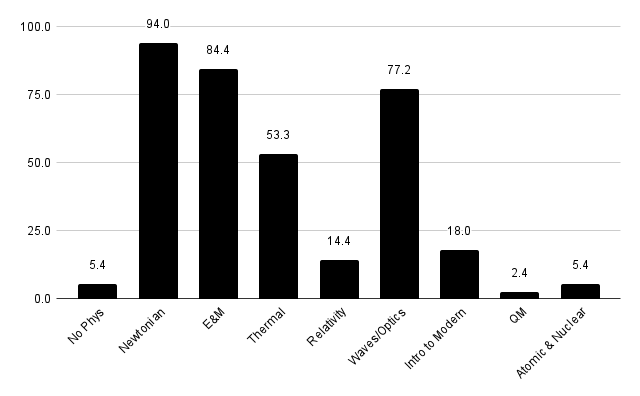}
    \caption{Required physics topics taught prior to enrollment in modern physics}
    \label{fig:enter-label}
\end{figure*}

\textit{Pre- and co-requisites requirements:}\\
The analysis of the course descriptions, concluded that the majority of students enrolled in Modern Physics have already taken Calculus II (37.1\%), followed by the next largest percentage of students having already taken Calculus III (22.8\%) as shown in Figure 4. As for physics backgrounds, students most commonly have previously been enrolled in courses that have introduced students to Newtonian physics (94\%), Electricity \& Magnetism (84.4\%), and Waves/Optics (77.2\%) (see Figure 5). 

\textit{Intended major and year of students enrolled:}\\
\begin{figure}
    \centering
    \includegraphics[width=200 pt, height=100 pt]{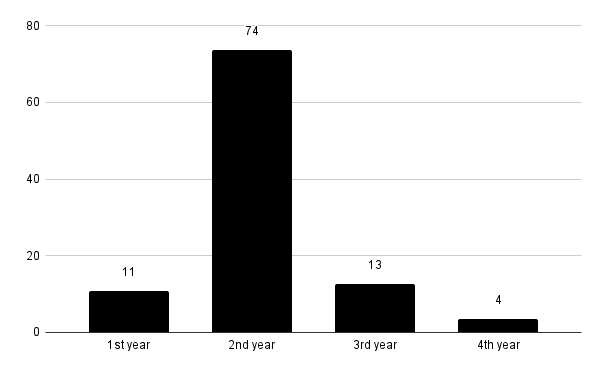}
    \caption{Intended year of enrollment}
    \label{fig:enter-label}
    \includegraphics[width=200 pt, height=100 pt]{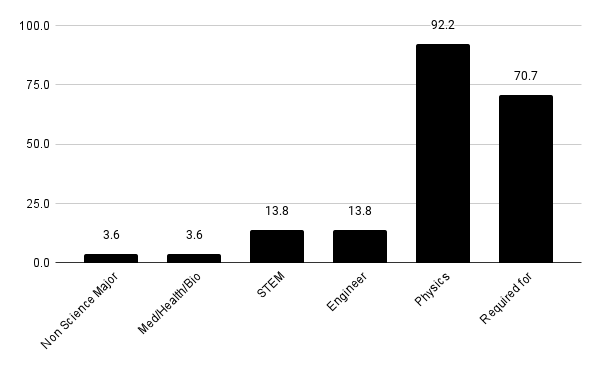}
    \caption{Intended audience of the course and if intended for physics majors, if it is required for physics majors to graduate}
    \label{fig:enter-label}
\end{figure}
74\% of institutions expect students to enroll in Modern Physics during their second year (Figure 6). It should be noted that the 4\% of institutions aiming modern physics courses at 4th year students were teaching traditional quantum mechanics courses but titled their course as modern physics, making them an outlier in the data set.  While 92.2\% of courses were intended for physics majors, only 70.7\%  of those were explicitly required for physics majors to graduate (Figure 7).

\textit{Pedagogical approach and grading scheme:}\\
\begin{figure}
    \centering
    \includegraphics[width=200 pt, height=100 pt]{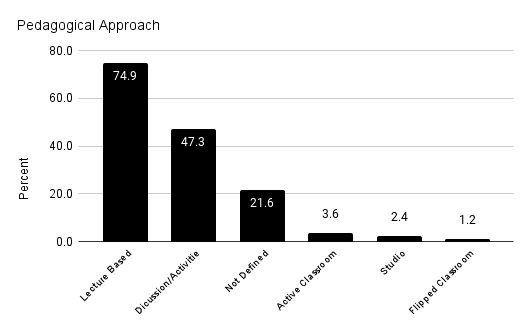}
    \caption{Pedagogical approach utilized}
    \label{fig:enter-label}
\end{figure}
Of the 167 syllabi, 21.6\% did not define a pedagogical approach (Figure 8). 74.9\% were lecture based, with 47.3\% having additional discussions, activities, or participation components built into the lecture or designated for a different time. 3.6\% used an active classroom format, 2.4\% used studio style and 1.2\% used a flipped classroom approach. Some courses that used active, studio, or flipped classroom approaches also had designated lecture times built in as well and were included in the 74.9\% of lecture based courses.

\begin{figure}
    \centering
    \includegraphics[width=200 pt, height=100 pt]{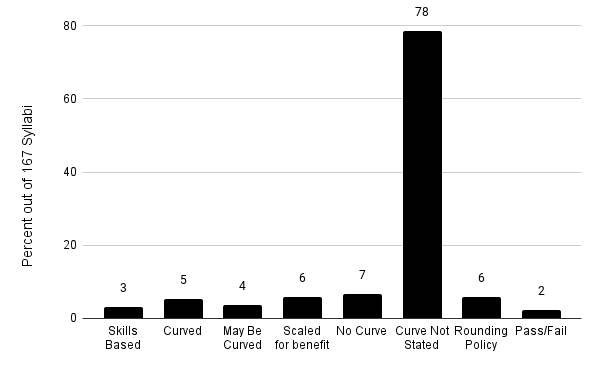}
    \caption{Grading scheme utilized}
    \label{fig:enter-label}
\end{figure}

78\% of syllabi did not state if there would or would not be a curve (Figure 9). 7\% stated there would be no curve, while 6\% stated the grade would only curve for the students benefit. 4\% stated there may be a curve without indicating if student grades would increase or decrease as a result of said curve. 5\% explicitly stated there would be a curve, but did not state that students would necessarily benefit from it. 3\% explicitly used a skills based, or absolute scale, stating all students could hypothetically receive an A. 6\% of syllabi has a policy on what grade would result if a students grade was on a letter grade boundary. 2\% of courses were using a pass/fail grading system. 

Of the 167 syllabi, 3\% explicitly state exams would not be used as an assessment tool,  7\% syllabi did not state if there were or were not exams, while 90\% explicitly stated there would be exams (Figure 10). Of the 90\% syllabi with exams, 50\% had a cumulative exam, 10\% had only noncumulative exams, and 40\% did not state if exams would be cumulative or not (Figure 11). 

\begin{figure}
    \centering
    \includegraphics[width=200 pt, height=100 pt]{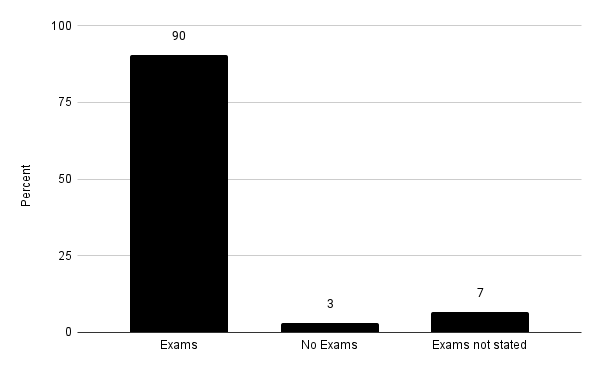}
    \caption{Use of exams as assessment tool in modern physics courses}
    \label{fig:enter-label}
    \includegraphics[width=200 pt, height=100 pt]{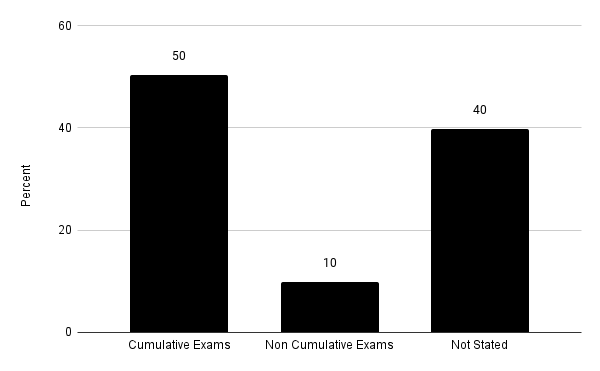}
    \caption{Use of cumulative and noncumulative exams}
    \label{fig:enter-label}
\end{figure}

\textit{Policies:}\\
\begin{figure*}
    \centering
    \includegraphics[width=400 pt, height=300 pt]{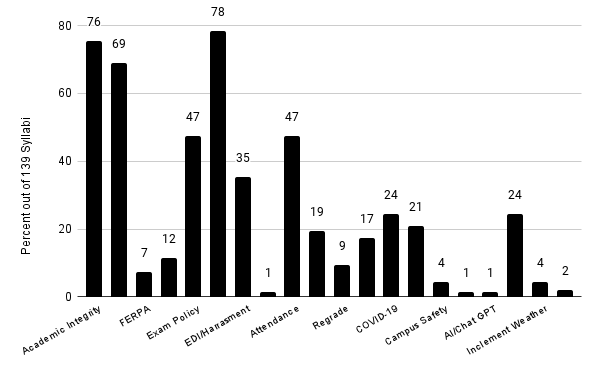}
    \caption{Policies listed in modern physics syllabi}
    \label{fig:enter-label}
\end{figure*}
At least one policy was listed in 83.2\% ($n=139$) of the syllabi. The most common policy, with 78\% of the 139 syllabi, was about late or makeup work. As shown in Figure 12, 76\% had a statement on academic integrity and 69\% had a policy or university policy listed on ADA and Accommodations.

\textit{Topic modeling using NLP:}\\
Running \texttt{BERTopic} on the same corpus 20 times produced between 97 and 105 topics, due to the stochastic nature of the underlying algorithm. For each topic, the algorithm provides the 10 words most associated with the topic and a measure of their relative weight; we mapped sentence fragments back to their parent syllabi to compute
the frequency with which each topic occured in the original corpus. In table \ref{tab:Bertopic results}, we show the ten topics that appeared most commonly in the syllabi from a typical run with BERTopic---these were robust across runs---together with the ten words most associated with each topic. We also provide a prototypical example sentence close to the center of the topic cluster, automatically generated by the algorithm Finally, we also provide a brief researcher generated interpretation of the topic. 

\begin{table*}
\begin{tabular*}{1\textwidth}{@{\extracolsep{\fill}}|c|>{\centering}p{2.5in}|>{\centering}p{2.5in}|>{\centering}p{1in}|}
\hline 
Frequency & Ten most relevant words & Prototypical sentence fragment & Interpretation\tabularnewline
\hline 
\hline 
121 (73\%) & {\scriptsize{}grading, grades, grade, graded, grader, exam, assignment,
scores, quizzes, calculated} & {\scriptsize{}``grading policy grade components your semester average
will be determined as follows...''} & How grades are calculated\tabularnewline
\hline 
120 (72\%) & {\scriptsize{}syllabus, instructor, lecturer, edu, prof, professor,
college, prerequisites, curriculum, astronomy} & {\scriptsize{}---} & Common words that appear in syllabi\tabularnewline
\hline 
100 (60\%) & {\scriptsize{}disabilities, disability, disabilityservices, accommodations,
accessibility, accommodation, rehabilitation, handicapped, eligibility,
eligible} & {\scriptsize{}``americans with disabilities act students with disabilities
needing academic accommodations should...''} & Academic accomodations\tabularnewline
\hline 
97 (58\%) & {\scriptsize{}textbooks, textbook, books, fundamentals, texts, isbns,
library, ebook, book, isbn} & {\scriptsize{}``required and recommended materials text book any
calc based text with modern physics physics for scientists and engineers...''} & Suggested textbooks\tabularnewline
\hline 
94 (58\%) & {\scriptsize{}lectures, lecture, study, textbooks, textbook, reading,
texts, readings, courseworks, notes} & {\scriptsize{}``be diligent about the reading assignments be on time
to class and turn in your completed homework when you arrive''} & Lecture component\tabularnewline
\hline 
91 (55\%) & {\scriptsize{}exams, exam, examinations, examination, midterm, midterms,
quizzes, schedule, final, testing} & {\scriptsize{}``the midterm exams will be held in rooms to be announced
in class and will take place during the scheduled quiz time see above''} & Exam policies\tabularnewline
\hline 
85 (51\%) & {\scriptsize{}homeworks, tutors, lateness, tutoring, deadline, homework,
late, credit, overdue, penalized} & {\scriptsize{}``late homeworks will be accepted with a credit penalty
through friday at the beginning of the class''} & Homework policies\tabularnewline
\hline 
71 (43\%) & {\scriptsize{}calculators, calculator, calculate, formulas, calculations,
equations, calculation, numerical, numerically, formulae} & {\scriptsize{}``calculators are allowed and a formula sheet together
with physical constants can be used''} & Policy on use of calculators\tabularnewline
\hline 
67 (40\%) & {\scriptsize{}misconduct, disciplinary, sanction, integrity, expulsion,
consequences, violates, student, violations, academic} & {\scriptsize{}``academic misconduct is a violation of the {[}redacted{]}
student code of conduct subject to a maximum sanction of disciplinary
suspension or expulsion as well as a grade penalty in the course''} & Academic misconduct policies\tabularnewline
\hline 
66 (39\%) & {\scriptsize{}instruction, lecturers, learning, scholarship, pursuing,
literacy, learn, struggling, design, barriers} & {\scriptsize{}``there are many ways to get help in this course and
we hope you do contact any member of the instructional team if you
feel unsure about the material and worry about your grade''} & Support mechanisms\tabularnewline
\hline 
\end{tabular*}

\caption{\label{tab:Bertopic results}Ten most frequent topics identified by
BERTopic, showing for each topic: its frequency in the corpus of syllabi,
the top ten words associated with the topic, a prototypical sentence
and a brief researcher-generated interpretation.}
\end{table*}

Most topics identified concern expected content of syllabi including
various components and policies: The sixth topic in table \ref{tab:Bertopic results},
for example, concerns exams, and occurs in 91/169 documents, a number
that matches almost exactly with the human-coded results in Figure
10. Importantly, however, the NLP algorithm is \emph{not} able to
resolve small number of examples of a ``no exam'' policy, likely
because these are very rare in the dataset. 

Not all topics generated were useful for our analysis; the second, for which we do not provide a protoypical sentence, appears to correspond to common words appearing in syllabi; such topics that appeared to neither be policy or content were assigned to a "not used" category. The remainder of the topics were divided into "policy" and "content"
topics and used to cross-validate our human coding process.

\begin{figure*}
\includegraphics{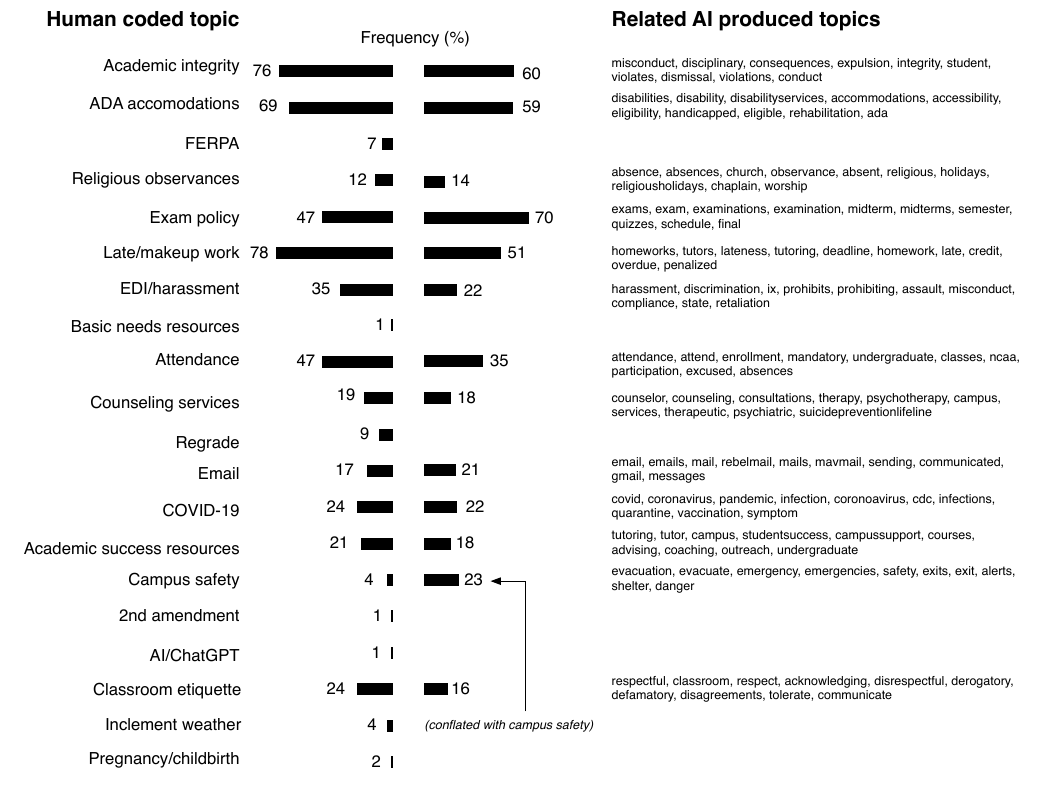}

\caption{\label{fig:Comparison-of-policy}Comparison of policy topics identified
by human coding (left) and NLP BERTopic coding with associated words
(right). }

\end{figure*}

We began by comparing the emergent topics from the NLP analysis to
the policies identified in Fig. 12. For each human coded topic, we
identified similar NLP topics by looking for similar words. The results
of this analysis are displayed in Fig. \ref{fig:Comparison-of-policy},
which places the frequency distributions for human codes alongside
that computed for the largest closely related topic identified by
the BERTopic algorithm. The relative frequencies are strikingly similar
for many topics, providing substantial evidence validating the codes
chosen and coding procedure used. 

\begin{table*}
\begin{tabular*}{1\textwidth}{@{\extracolsep{\fill}}|c|>{\centering}p{2in}|>{\centering}p{4in}|}
\hline 
Frequency & Ten most relevant words & Prototypical sentence fragment\tabularnewline
\hline 
\hline 
{\small{}70 (41\%)} & {\scriptsize{}misconduct, disciplinary, consequences, expulsion, integrity,
student, violates, dismissal, violations, conduct} & {\scriptsize{}students must recognize that failure to follow the rules
and guidelines established in the university s code of student conduct
and this syllabus may constitute academic misconduct...}\tabularnewline
\hline 
{\small{}40 (24\%)} & {\scriptsize{}integrity, academicintegrity, scholarly, academic, scholars,
institution, scholarship, faculty, trust, excellence} & {\scriptsize{}as described in the {[}Redacted{]} academic integrity
is the basic guiding principle for all academic activity at penn state
university allowing the pursuit of scholarly activity in an open honest
and responsible manner}\tabularnewline
\hline 
{\small{}27 (16\%)} & {\scriptsize{}integrity, academic, honesty, ethics, ethical, education,
scholarly, academichonesty, umbc, informational } & {\scriptsize{}academic honesty {[}Redacted{]} has a comprehensive
academic honesty policy document a culture of honesty which is available
from office of the vice president for instruction at ...}\tabularnewline
\hline 
26 (15\%) & {\scriptsize{}honor, integrity, pledge, academic, university, upholding,
studenthealth, academics, uphold, acceptance} & {\scriptsize{}the honor code reads as follows to promote a stronger
sense of mutual responsibility respect trust and fairness among all
members of the {[}Redacted{]} community and with the desire for greater
academic and personal achievement ...}\tabularnewline
\hline 
\end{tabular*}

\caption{\label{tab:Misconduct}BERTopic generated topics associated with the
human coded topic ``Academic integrity''}
\end{table*}

In some cases, the NLP algorithm identified more than one topic that
appears to be related to the human code. For example, the human code
``Academic integrity'' appears to be related to four topics displayed
in Table \ref{tab:Misconduct}. The topics identified by NLP may nonetheless
capture different aspects of academic integrity: The first may be
associated with academic misconduct and violations or consequences;
the last seems to be associated with honor codes. Researchers encountering
a similar phenomenon could choose a number of approaches: they might
choose to further investigate the differences between topics either
through human reading or by computing similarity scores; or, they
may choose to collapse selected topics. Here we collapse all topics
associated with a human code into a single topic and use the size
of the combined topic for the frequency count.

As for the simple example on exam policies presented earlier, topics
that were identified through human coding that are rare in the dataset
are not resolved by the NLP algorithm. No topics were identified by
BERTopic that are related to any of the FERPA, Basic needs assistance,
regrading policy, the 2nd amendment, AI/ChatGPT (the researchers noted
this omission with amusement) or Pregnancy/Childbirth codes. These
elements are certainly in the text, but are too infrequent to survive
the clustering process. Some of these topics, such as the latter,
are closely related to issues of equity, and hence we observe an important
possible source of bias if NLP analysis were to be relied on exclusively. 

The NLP analysis \emph{did} identify some policy topics that appear
to be distinct from those chosen in the human coding process. We display
these in table \ref{tab:AlternativeTopics}, together with a brief
researcher-generated interpretation. A third of syllabi mention policies
related to athletics or extracurricular activities, which was not
coded for in the human analysis. Other topics are, arguably, related
to human coded topics: incomplete grades and withdrawals are related
to grading policies overall, and copyright and citations to academic
integrity. Nonetheless, it is interesting to note that some syllabi
explicitly mention these distinctly. While we did not do so here,
NLP generated topics could serve as a basis for follow up analysis. 

\begin{table*}
\begin{tabular*}{1\textwidth}{@{\extracolsep{\fill}}|c|>{\centering}p{2in}|>{\centering}p{3in}|>{\centering}p{1in}|}
\hline 
Frequency & Ten most relevant words & Prototypical sentence fragment & Interpretation\tabularnewline
\hline 
\hline 
59 (35\%) & {\scriptsize{}extracurricular, curricular, activities, intercollegiate,
athletics, athletic, excused, activity, competitions, accrued} & {\scriptsize{}for purposes of definition extracurricular activities
may include but are not limited to academic recruitment activities
competitive intercollegiate athletics fine arts activities liberal
arts competitions science and engineering competitions and any other
event ...} & Extracurricular and Athletic activities\tabularnewline
\hline 
33 (20\%) & {\scriptsize{}incompletes, incomplete, grade, semesters, gpa, deadline,
certifiable, nullified, requirements, absence} & {\scriptsize{}incompletes you may be assigned an incomplete for the
course in accordance with the uga regulations provided all of the
following applies you received a non failing grade in labs you received
a non failing grade on at least one exam no violation of the academic
honesty policy took place during the course of the semester} & Incomplete grades\tabularnewline
\hline 
22 (13\%) & {\scriptsize{}withdrawn, withdrawing, withdrawal, withdraw, {[}Redacted{]},
withdrawals, {[}Redacted{]}, {[}Redacted{]}, deadline, deadlines} & {\scriptsize{}for medical withdrawals requests to college to be dropped
from a class after the deadline for withdraw has passed the withdraw
pass wp or withdraw fail wf grade will usually be determined by the
pro rated grade...} & Withdrawals\tabularnewline
\hline 
17 (10\%) & {\scriptsize{}copyright, copyrighted, infringement, copying, license,
documents, infringe, copied, violates, prohibited} & {\scriptsize{}copying displaying reproducing or distributing copyrighted
works may infringe the copyright owner's rights and such infringement
is subject to appropriate disciplinary action as well as criminal
penalties provided by federal law} & Copyright\tabularnewline
\hline 
12 (7\%) & {\scriptsize{}citations, citing, bibliography, cite, citation, cited,
references, researching, wikipedia, guides} & {\scriptsize{}citation is commonly done use a style manual which provides
guidance on how to format the information for a citation such as title
author pages and date as well as formatting and grammar specifics} & Citations\tabularnewline
\hline 
\end{tabular*}

\caption{\label{tab:AlternativeTopics}Topics identified by BERTopic that are
distinct from those identified by human coding.}
\end{table*}

\begin{figure*}
\includegraphics{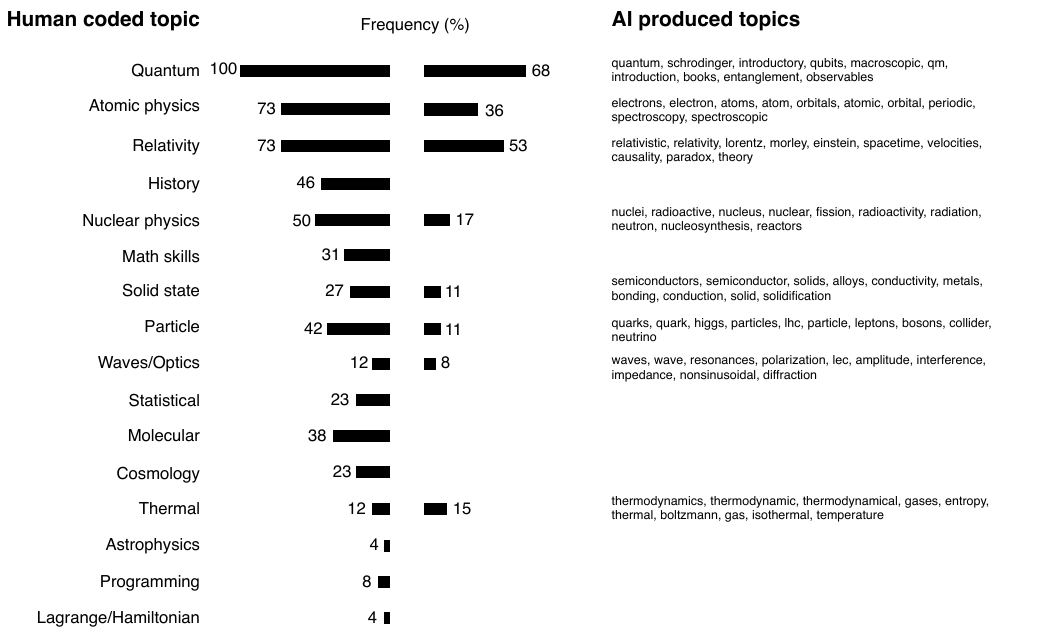}

\caption{\label{fig:ContentTopics}Comparison of content topics identified
by human coding (left) and NLP BERTopic coding with associated words
(right). }
\end{figure*}

We identified fifteen emergent topics that correspond to course content
as shown in Fig. \ref{fig:ContentTopics}. By comparing these with
codes used for the manual coding process, each NLP generated content
topic was assigned to a human coded topic. Only one topic was not
assigned that occurred in 21 documents and was associated with the
words \emph{``circuits, electrodynamics, electromagnetism, electromagnetic,
conductors, faraday, electricity, electric, electrostatics, currents''}.
As can be seen in Fig. \ref{fig:ContentTopics}, while many of the
content topics were also identified by the NLP analysis, the frequency
estimates are considerably poorer than those from the policy analysis.
This is likely due to the language model used by BERTopic, that may
not properly capture physics and math terminology. 

\section{Discussion}
With quantum being the most commonly taught topic, and 70.7\% of modern physics courses intended for physics students being a graduation requirement, it can be concluded that modern physics is students' first introduction to quantum concepts. This conclusion echoes the speculation made by Vokos et al. \cite{Vokos2000} and Singh et al. \cite{Singh2006}. However, because the majority of students enrolled in modern physics have not yet seen linear algebra or differential equations, the modern physics course must only be an introduction to quantum mechanics. Without understanding of the more advanced mathematics courses, students are not yet ready to tackle problems in commonly used undergraduate textbooks such as Griffiths \cite{Griffiths} or McIntyre \cite{McIntyre}. Therefore, in order for students to fully understand the Schr{\"o}dinger equation and be able to solve problems related to it with little to no assistance, they will need another course later in their undergraduate career on quantum mechanics. Without having seen linear algebra prior to enrollment, it would be interesting to determine if any modern physics courses are using a spins first approach to quantum concepts. 

Physics educators may want to teach students more about quantum mechanics earlier on in their undergraduate career, but the reality is that the majority of students come to university under prepared. Fewer than one in four American 12th graders performed proficiently in math on the National Assessment of Educational Progress in 2019 \cite{Nationsreportcard}. The uniqueness of individual institutions' student bodies must be considered when designing a program and determining what level of introduction to quantum students are ready for, particularly concerning math preparedness.

It can also be drawn from these results, that research intensive programs have largely converged on a certain set of topics for the modern physics course without having a community conversation or consensus. A syllabi analysis of smaller, teaching focused institutions modern physics courses could reveal interesting results on whether or not they converged on the same set of topics as the research intensive institutions. If the topics were to be different, this could be the start of a community conversation or consensus to increase equal access to education regardless of institution type. 

Introductory physics courses, such as classical mechanics and electricity and magnetism, have undergone significant changes to increase the level of interactivity and decrease the amount of lecturing. However, from the results in this study, we can see the level of interactivity has significantly dropped in the modern physics course. There is a significant opportunity for pedagogical innovation within the modern physics course that is not currently being leveraged. With modern physics being a gateway course to the physics major, physics educators should strive for this course to be inviting. This course could offer a potential area to experiment with different pedagogical approaches and grading schema from the traditional approaches. The level of programming opportunities mentioned within the syllabi was low (5\%), and could be another opportunity for introduction to computation that is important for physics careers or research students may pursue. 

In this study, topic modeling was performed, an NLP task, as a methodology for cross-validation of a human-centered coding process. This is possible due to the tractable size of the authors' corpus for human analysis, but offers valuable insight into the promise, potential applications and limitations of NLP methods in qualitative work. With modest effort on the part of the researcher, NLP methods provide a broad perspective of the content of a corpus. Like others \cite{PhysRevPhysEducRes.19.020123}, we find that transformer-based algorithms generated more immediately interpretable results than earlier
methods with little tuning of hyperparameters. 

In the present case, most of the human identified codes were also separately identified by the BERTopic algorithm, suggesting that generated topics are a good starting point for qualitative analysis, at least for the kind of curricular data discussed here. By design, to facilitate validation, we conducted the NLP analysis entirely separately from the human coding process. Nonetheless, we concur with others \cite{PhysRevPhysEducRes.19.020123,PhysRevPhysEducRes.16.010142,odden2024using} that such analyses can be synergistic, and believe that iteration of human and NLP analysis would be beneficial in other contexts. 

A particular focus for human intervention is that the NLP analysis does not capture \emph{rare} but potentially interesting features of the dataset. In this sample, for example, the NLP analysis was not able to capture the fact that 2\% of the syllabi referred to pregnancy, 1\% referred to AI/ChatGPT, and 3\% referred to Lagrangian/Hamiltonian mechanics; neglecting the uncommon has important implications for equity. Further, the frequencies with which topics were identified
showed similar overall trends between both methods. We noticed a broad tendency of the NLP algorithm to \emph{under}-count topics, particularly for the topics associated with the course content. The undercounting may reflect biases of the training set of the BERT embedding used; language models specifically trained on physics and math texts may therefore considerably enhance the accuracy of the results. While not used here, LLMs that utilize larger models may also add utility, particularly because LLMs could be used to generate descriptions of clusters from the corpus, which may further improve the interpretability of topics.

\section{Conclusion}
Quantum concepts are most commonly first introduced to students in their modern physics courses, making the course a pivotal experience for the physics major. With most students only having completed calculus II at the time of enrollment, students will require another course on quantum mechanics in order to solve the Schr{\"o}dinger equation on their own. The modern physics course also opens opportunities for institutions to work together to lessen disparities in educational access regardless of institution type, ensuring all students are offered a comprehensive education. Additionally, there is opportunity for instructors to implement more interactive pedagogical approaches and innovative grading schemes, rather than falling back on the traditional approaches. NLP as a methodology for cross-validation of human-centered coding has shown promise, as they were often in aggreeance but the human-centered coding revealed rare or interesting cases the NLP missed, and the NLP analysis did identify aspects that appear to be distinct or nuanced from those chosen in the human coding process. 

\clearpage

\appendix

\section{Codes for content taught}

\center
\begin{longtable*}{ |c|c|c|  }
 \hline
 \multicolumn{3}{|c|}{Content Codes} \\
 \hline
    Thermal& Relativity& Quantum\\
 \hline
    Thermal equilibrium& Special relativity& Schr{\"o}dinger\\
    Entropy& General Relativity& Schr{\"o}dinger equation\\
    Heat& Spacetime& Photoelectric effect\\
    p-V diagrams& Invariants& Wave-particle duality\\
    Ideal gas law& Frame Transforms& Operators\\
    Kinetic theory& Lorentz& Eigenvalues/vectors\\
    Pressure& 4-vector& Tunneling/reflection\\
    Temperature& Metric tensor& Stern-Gerlach experiment\\
    Temperature& Minkowski& Dirac notation\\
    Heat capacity& Michelson-Morley experiment& States\\
    Specific heat& Time dilation& Quantum measurement\\
    Carnot cycle & Length contraction& Expectation value\\
    Bernoulli's equation& Energy-momentum& Uncertainty\\
    Pascal's principle& Classical relativity& Superposition\\
    Archimedes' principle& Einstein's postulates& Mixed states\\
     & Twin paradox& Quantization\\
     & Relativistic dynamics& Fermi's golden rule\\
     & Relativistic energy& Photons\\
     & Relativistic momentum& Pauli's exclusion principle\\
     & Mass-energy equivalence & Square well\\
     & & Identical particles\\
     & & Matter waves \\
     & & Frank Hertz experiment\\
     & & Wave mechanics\\
     & & Wave functions\\
     & & Wave properties of particles\\
     & & Particle properties of waves\\
     & & de Broglie hypothesis\\
     & & Quantum theory of light\\
     & & Blackbody radiation\\
     & & Planck's postulate\\  
 \hline
    Atomic& Nuclear& Molecular\\
 \hline
    Atomic& Nucleus& Molecules\\
    Atom& Nuclear Atom& Bonds\\
    Bohr model& Fission& Molecular spectroscopy\\
    Thomson/plum pudding model& fusion& Quantum theory of molescules\\
    Rutherford model/experiment& Decay& Chemical bonding\\
    Zeeman effect& Radioactivity& Vibrational and rotational energies of molecules\\
    Hydrogen& Strong interaction& \\
    Many electron atoms& Weak interaction& \\
    Spectra& Alpha decay& \\
    Emission/absorption& Beta decay& \\
    Periodic table& Gamma decay& \\
    Scanning tunneling microscopy& Nuclear force& \\
    & Electron capture& \\
 \hline
    Solid State& Statistical& Cosmology\\
 \hline
    Solids& Bose-Einstein& Chronology of universe\\
    Semiconductor& Fermi-Dirac& Big bang theory\\
    Superconductivity& Quantum statistics& Evolution of universe\\
    Band structure& Classical statistics& Structure of universe\\
    pn-junction& Maxwell-Boltzmann& Cosmic Microwave Background\\
    Condensed matter& Classical gas& \\
    Crystal structure& Quantum gas& \\
 \hline
    Programming Skills& Math Skills& History\\
 \hline
    Numerical Investigation & Operators& Michelson-Morley Experiment\\
    Mathematica& Eigenvalues/vectors& Photoelectric effect\\
    Igor Pro& Dirac notation& Stern-Gerlach experiment\\
    Python& Simple harmonic oscillator& Frank Hertz experiment\\
    Numerical project& Simple harmonic motion& Compton effect/scattering\\
    Computational Project& Fourier analysis& Bohr model\\
    & Matrices& Thomson model\\
    & Complex algebra& Rutherford model\\
    & Hilbert space& Milikan Oil Drop experiment\\
    & Mathematical description of waves& de Broglie hypothesis\\
    & Normalization& Einstein's postulates\\
    & Complex notation& Blackbody radiation\\
    & probability& Double slit experiment\\
    & Expectation value& Classical vs quantum measurement\\
    & Spherical coordinates& Planck's postulate\\
    & Radial equation& Michelson interferometer\\
    & Math review& Origins of quantum mechanics\\
    & Symmetries& Early quantum theory\\
    & & Higgs boson\\
    & & birth of quantum mechanics\\
    & & Quantum paradoxes\\
 \hline
    Particle& Waves/optics& Astrophysics\\
 \hline
    Standard model& Electromagnetic waves& Stars\\
    Fundamental interactions& Resonance& Celestial bodies\\
    Quark model& Oscillations& Newtonian gravitation\\
    Bosons& Interference& Kepler's laws\\
    Fermions& Diffraction& Orbits\\
    Neutrinos& Sound& Spectroscopy in astronomy\\
    Higgs boson& Doppler effect& \\
    & Reflection& \\
    & Snell's law& \\
    & Mirrors& \\
    & Lenses& \\
    & Polarization& \\
    & Classical waves& \\
 \hline
    Lagrangian/Hamiltonian Mechanics & Other uncoded topics& \\
 \hline
    & Quantized electromagnetic fields& \\
    & Quantum electrodynamics& \\
    & Quantum chromodynamics& \\
    & Quantum field theory& \\
 \hline
\caption{\label{tab:Topicscoding}Codes for content taught.}
\end{longtable*}

\section{Codes for pedagogy used}

\begin{longtable*}{ |c|c|c|  }
 \hline
 \multicolumn{3}{|c|}{Pedagogy Codes} \\
 \hline
    Lecture based& Activities accompanying lectures& Studio based\\
 \hline
    Any instance of & Discussions/activities/recitations/tutorials & Any instance of \\
    the word lecture occurring & supplementing lecture based class& the word "studio"\\
    & Clicker questions in lecture& \\
    & In class assignments& \\
    & Students present solutions to class& \\
    & Class participation in lecture required& \\
 \hline
    Active Classroom& Reverse Classroom& Not Defined \\
 \hline
    Any instance of the word& Any instance of the word & Pedagogy not stated,\\
     "active" to describe classroom environment& "reversed" or "flipped"& requirements for other\\
    Uses lecture/& to describe classroom& pedagogy codes not met\\
    class time exclusively & & \\
    for discussions or activities& & \\
 \hline
\caption{\label{tab:pedagogycoding}Codes for pedagogy used.}
 \end{longtable*}

\section{Codes for grading scheme used}

 \begin{longtable*}{ |c|c|c|  }
 \hline
 \multicolumn{3}{|c|}{Grading Scheme Codes} \\
 \hline
    Skills based& Curved& Curve not stated\\
 \hline
    Everyone can get an A& Explicitly states there & No clear statement\\
    Students not in competition& will be a curve& whether grade will be\\
    No desired bell curve& Letter grade for percentage& curved or not\\
    or grade distribution& score not determined till& \\
    & course complete& \\
    & Sliding scale that does not& \\
    & specify if it will benefit& \\
    & students or not& \\
    & A tentative scale is given& \\
 \hline
    May  be curved& No curving& Scaled only for students benefit\\
 \hline
    Curve may or may not& Explicitly states course will& Curve/Scale explicitly\\
    be applied depending on& not be curved (but did not& stated will only\\
    distribution& state everyone can get an A)& increase students grades\\
    Nothing explicitly stating if& Absolute scale used& \\
    this curve will benefit or& & \\
    hurt students grades& &\\
 \hline
    Rounding policy& Pass/Fail& \\
 \hline
   States how students grades& &\\
   will be rounded if on letter& & \\
   grade boundary& & \\
 \hline
\caption{\label{tab:gradingcoding}Codes for grading scheme used.}
 \end{longtable*}

\section{Codes for policies listed}
\begin{longtable*}{ |c|c|c|  }
 \hline
 \multicolumn{3}{|c|}{Policy Codes} \\
 \hline
    Are policies included?& Academic Integrity& Disabilities/Accommodations\\
 \hline
    Codes for if any policy& Any statement about& Americans with Disabilities\\
    was listed& cheating, plagiarism, or& Act\\
    & honor code& Accomodations\\
 \hline
    FERPA& Religious Observances& Exam Policies\\
 \hline
    Any statement about FERPA& & Materials allowed on exams\\
    or student privacy or records& & \\
    Policy on recording lectures& & \\
 \hline
 Late/Makeup work& EDI or harassment& Basic needs resources\\
 \hline
 Exam makeups& Title IX statement& Food\\
 Extensions& Sexual harassment statement& Shelter\\
 & EDI statement& Sleep\\
 & & Nutrition\\
 \hline
 Attendance& Counseling services& Regrade policies\\
 \hline
 Is there an attendance& Intuition counseling center& How regrades will be handled\\
 policy listed?& or mental health services& and timeline allocated\\
 & listed& to request a regrade\\
 \hline
 Email policy& COVID-19 policy& Academic success resources\\
 \hline
 How professor prefers& Mask policy& Time management coaches\\
 to be contacted& COVID reporting policy& Writing centers\\
 How to email professor& & Tutors\\
 \hline
 Campus Safety& 2nd Amendment& AI/ChatGPT \\
 \hline
 Evacuation plans& Open carry policies& \\
 \hline
 Class etiquette& Weather& Pregnancy/Childbirth\\
 \hline
 Classroom expectations on& Statement on cases& \\
 behavior& of inclement weather& \\
 Cell phone, laptop, or& & \\
 electronics usage in class& & \\
 Civility statement& & \\
\hline
\caption{\label{tab:policycoding}Codes for policies included.}
 \end{longtable*}

\bibliography{bib}

\end{document}